\documentclass[twocolumn,showpacs,amsmath,amssymb,pre,aps]{revtex4}   
\usepackage{graphicx}
\usepackage{dcolumn}
\usepackage{bm}
\usepackage{rotating}

\begin{document}

\title{Negative differential resistivity in superconductors with periodic
arrays of pinning sites}

\author{Vyacheslav R. Misko$^{1,2,3}$, Sergey Savel'ev$^{1,4}$, Alexander L. Rakhmanov$^{1,5}$,
and Franco Nori$^{1,2}$} \affiliation{$^1$ Frontier Research
System, The Institute of Physical and Chemical Research (RIKEN),
Wako-shi, Saitama, 351-0198, Japan} \affiliation{$^2$ Center for
Theoretical Physics, Department of Physics, University of
Michigan, Ann Arbor, MI 48109-1040, USA} \affiliation{$^3$
Department of Physics, University of Antwerpen (CGB), B-2020
Antwerpen, Belgium} \affiliation{$^4$ Department of Physics,
Loughborough University, Loughborough LE11 3TU, United Kingdom}
\affiliation{$^5$ Institute for Theoretical and Applied
Electrodynamics Russian Academy of Sciences, 125412 Moscow,
Russia}

\date{\today}

\begin{abstract}
We study theoretically the effects of heating on the magnetic flux
moving in superconductors with a periodic array of pinning sites
(PAPS). The voltage-current characteristic (\textit{VI}-curve) of
superconductors with a PAPS includes a region with
\textit{negative differential resistivity} (NDR) of S-type (i.e.,
S-shaped \textit{VI}-curve), while the heating of the
superconductor by moving flux lines produces NDR of N-type (i.e.,
with an N-shaped \textit{VI}-curve). We analyze the instability of
the uniform flux flow corresponding to different parts of the {\it
VI}-curve with NDR. Especially, we focus on the appearance of the
filamentary instability that corresponds to an S-type NDR, which
is extremely unusual for superconductors. We argue that the
simultaneous existence of NDR of \textit{both} N- and S-type gives
rise to the appearance of self-organized two-dimensional dynamical
structures in the flux flow mode. We study the effect of the
pinning site positional disorder on the NDR and show that moderate
disorder does not change the predicted results, while strong
disorder completely suppresses the S-type NDR.
\end{abstract}


\maketitle

\section{Introduction}

Negative Differential Resistivity (NDR) and Conductivity (NDC) can
be observed in various non-linear media. To illustrate the
counterintuitive nature of this phenomenon, let us consider a
force acting on a set of moving particles: NDC corresponds to a
\textit{lower} velocity of motion for these particles when the
force applied to them \textit{increases}. Two different types of
NDR can be observed in the voltage-current characteristics
(\textit{VI}-curves) of non-linear
media~\cite{sch,sha,MR,GM,GMR,cap,mikhailovskii}. NDR of S-type is
characterized by the existence of \textit{three} different values
of the current $I$ corresponding to a \textit{single} value of the
voltage $V$. The corresponding \textit{VI}-curve is S-shaped. A
\textit{VI}-curve with \textit{three} different values of the
voltage for a \textit{single} value of the current is referred to
as NDR of N-type. The corresponding \textit{VI}-curve is N-shaped.
NDR (or NDC) is commonly observed in semiconductors, plasmas,
superconductors and is used in many non-linear devices (see, e.g.
Refs.~\onlinecite{sch,sha,MR,GM,GMR,cap,mikhailovskii}). In
particular, semiconductors with NDR are the basic elements of
Gunn-effect diodes and $pnpn$-junctions~\cite{sch,sha}.
\textit{VI}-curves with NDR can only be observed under specific
conditions. For example, to study N-type NDR one has to include
the corresponding sample in an electric circuit with fixed voltage
$V$. Vice versa, to observe S-type NDR the sample should be
included in a circuit with fixed current $I$. If these conditions
are not fulfilled, the uniform current flow becomes unstable, and
non-uniform self-organized structures (e.g., filaments and
overheated domains with higher or lower electric fields) arise in
the sample. Such structures are commonly observed in plasmas,
semiconductors, and superconductors (see, e.g.
Refs.~\onlinecite{sch,sha,MR,GM,GMR,cap,mikhailovskii}). Table~1
(using results from
Refs.~\onlinecite{sch,sha,MR,GM,GMR,cap,mikhailovskii,Hueb,LO,kun,doettinger,tokunaga1,tokunaga2}
compares NDR in different non-linear media. Note, the macroscopic
manifestations of NDR in different media could be rather similar
while the intrinsic physical mechanisms giving rise to NDR could
be different.

\begin{table*}
%
\begin{tabular}{|c|c|c|c|c|}
  \hline
   & {\bf Superconductors} & {\bf Semiconductors} & {\bf Plasmas} & {\bf Manganites} \\
  \hline
  {Carriers} & flux quanta & charge quanta:     & electrons & electrons, \\
                 &             & electrons or holes &           & holes      \\
\hline
  {Characteristic} & voltage-current     & current-voltage  & {\it IV} curve & {\it IV} curve \\
                  {curve}           & (\textit{VI}) curve & ({\it IV}) curve &                &                \\
\hline
  {Homogeneous} & homogeneous flux  & homogeneous  & homogeneous  & homogeneous  \\
   {state}                & and current flows & current flow & current flow & current flow \\
\hline
  {Origin of } & in/commensurate vortex   & non-linear electron & ionization & $-$ \\
  {S-shape NDR}               & dynamical phases in PAPS & transport           &            &     \\
   \hline
  {Origin of } & overheating, Cooper pair                                 & overheating, electron & heating & heating \cite{tokunaga1,tokunaga2} \\
  {N-shape NDR}               & tunnelling~\cite{Hueb}; vortex-core:                     & or hole tunnelling    &         &         \\
                          & shrinkage~\cite{LO} ($T \approx T_{c}$);                 &                       &         &         \\
                          & expansion~\cite{kun}, driven~\cite{doettinger} &                       &         &         \\
                          & ($T \ll T_{c}$) &                       &         &         \\
\hline
  {Filaments} & overheating, Cooper pair                                 & current filaments, & current filaments, & $-$ \\
              & tunnelling~\cite{Hueb}; vortex-core:                     & pinch-effect       & pinch-effect       &         \\
              & shrinkage~\cite{LO} ($T \approx T_{c}$);                 &                       &         &         \\
              & expansion~\cite{kun}, driven &                       &         &         \\
              & vortices~\cite{doettinger} ($T \ll T_{c}$) &                       &         &         \\
\hline
  {Domains} & vortex-induced higher $E$ & higher electric field & higher electric  & higher electric \\
            & field overheated domains  & overheated domains    & field overheated & field overheated \\
            &                           &                       & domains          & domains    \\
  \hline
\end{tabular}

\caption{\label{tab:table1} Comparison between non-uniform
non-equilibrium states in superconductors, semiconductors,
plasmas, and manganites with \textit{VI}-curves having Negative
Differential Resistivity (NDR) \cite{weprl}. Since $IV$-curves in
semiconductors map to $VI$-curves in superconductors, then NDC
(for semiconductors) maps into NDR for superconductors. Here,
N(S)-type shapes for semiconductors correspond to S(N)-type for
superconductors~\cite{GMR}. The Negative Differential Conductivity
(NDC) found in~\cite{kun} is analogous to the Gunn effect in
semiconductors, where electron-charge modulations lead to steps in
$j(E)$ in the NDC regime. }
\end{table*}

As a non-linear medium with NDR, we study a superconductor with
artificial pinning sites. The magnetic flux behavior in such
superconductors has attracted considerable attention due to the
possibility of constructing samples with enhanced pinning as well
as with novel and unusual voltage-current
characteristics~\cite{vvmdotprl,fnsc2003,rwdot,vvmfddot,penro,quasi}.
Present-day technology allows the fabrication superconductors with
well-defined periodic arrays of pinning sites (PAPS). Such
structures include many thousands of elements with controlled
microscopic pinning parameters. Increased interest on these
systems has arisen in recent years, and a number of intriguing
features related to PAPS has been revealed.

In Ref.~\onlinecite{fn1} the existence of several dynamical vortex
phases was predicted for square PAPS subjected to perpendicular
magnetic field, $B$, close to the first matching field
$B_\phi=\Phi_0/a^2$, where $\Phi_0$ is the magnetic flux quantum
and $a$ is the PAPS period. The geometry of the problem is shown
in Fig.~1 which is discussed in Sec. II. The effect of the dynamic
phases on the \textit{VI}-curve is illustrated in Fig.~2a. Let us
assume that the field $B$ is slightly higher than $B_\phi$ and the
number of the vortices in the sample $N_v$ is higher than the
number of the pinning sites $N_p$. Let us now slowly increase the
applied current $j$ in the sample. For very low current density
$j$ (phase I in Fig.~2) all vortices are pinned and their average
velocity $\bar{v}$ is zero. With increasing the current $j$,
interstitial vortices, $N_{\textrm{int}}=N_v-N_p$, start to move
and the velocity $\bar{v}$ becomes nonzero and grows with $j$
(phase II in Fig.~2). With further increasing the current $j$, the
driving force acting on a single vortex, $F_d=j\Phi_0/c$,
overcomes the pinning force, and a significant fraction of the
vortices start to move. This motion is uniform and very
disordered. Phase III corresponds to such vortex-flow mode. At
higher vortex velocities, the random motion of vortices becomes
more ordered. Some vortices become pinned in commensurate rows
while others move along vortex rows, which are incommensurate with
the underlying PAPS. Namely, when $j$ exceeds a threshold value,
only incommensurate vortex rows move and the vortex velocity
exhibits a significant drop with the increase of the driving force
(phase IV). Note that the phases III and IV have an analogy with
the NDR behavior of electron motion in \textit{semiconductors}
with the increase of the voltage. Namely, increasing the applied
force on the moving particles produces a lower velocity in them.
At high current densities, the driving force completely overcomes
the pinning force (phase V) and the curve $\bar{v}(j)$ tends to a
linear one.

The results obtained in Refs.~\onlinecite{fn1} and
\onlinecite{weprl} prove that the \textit{VI}-curves of
superconductors with a square PAPS have a part with NDR of the
S-type, since the electric field in the sample is related to the
average vortex velocity by the well-known relation
$E=-\bar{v}(j)B/c$. Such type of NDR is usual for
plasmas~\cite{cap,mikhailovskii} and semiconductors~\cite{sch,sha}
giving rise to important instabilities of the uniform current flow
known as pinch-effect and {\it filamentary instability}, when a
current flow breaks into filaments with lower and higher current
densities~\cite{sch,sha,MR,GM,GMR,cap,mikhailovskii}. In
superconductors we have only few examples of the S-type NDR in the
samples with the specific weak links~\cite{kun}.

The described dynamical phases disappear in the case of very
disordered pinning arrays and, consequently, the NDR of S-type in
the VI -curves vanishes. Under realistic experimental conditions,
the properties of the superconductor in the flux flow regime are
strongly affected by Joule heating since the current density $j$
necessary to overcome the pinning force is high~\cite{GMR}. An
increase of the sample temperature $T$ due to Joule heat, $jE$,
gives rise to a decrease of the pinning force, and the current
density can drop down with the growth of the electric field. As a
result, the \textit{VI}-curve with an NDR of N-type (red dashed
line in Fig.~2b) is commonly observed in superconductors for high
current density~\cite{GM,GMR}. The uniform state in samples with
NDR of N-type is also unstable~\cite{sch,sha}; and a propagating
resistive state boundary or the formation of resistive domains can
destroy the uniform flux flow mode in the
superconductor~\cite{GM,GMR,Hueb}. Increasing the pinning force
and/or decreasing the thermal coupling of the sample with its
environment (which decreases the cooling rate of the sample), one
can achieve a situation where the NDR of {\it both} N- and S-type
{\it simultaneously coexist} in the \textit{VI}-curve
(Fig.~2b)~\cite{weprl}. In this case, we predict remarkable flux
flow instabilities. Note also that the effect of thermal
fluctuations on the flux flow regime in the superconductors with
PAPS is somewhat analogous to the effect of positional disorder in
the pinning sites. This effect should be taken into account,
especially, at temperatures close to the critical temperature of
the the superconductor $T_c$.

The effect of Joule heating on the \textit{VI}-curve of
superconductors with PAPS was first outlined in
Ref.~\onlinecite{weprl}. This gives rise to the coexistence of
\textit{both} types of instabilities, which is very unusual since
non-linear devices are typically either N-type or N-type, but
\textit{not} both. Here we present a detailed analysis of this
problem. In addition, we study in detail the effect of disorder on
this phenomenon.

The paper is organized as follows. In Section II we formulate the
model of the flux motion taking into account the effect of Joule
heating. The heating manifests itself in thermal fluctuations of
the vortices and a variation of the superconductor parameters due
to the temperature increase. Both of these effects are included in
the model. In Section III the average velocity of the vortices,
$\bar{v}$, is found as a function of the applied current density
$j$. This dependence is obtained by means of the molecular
dynamics integration of the equations of motion presented in
Section II. As a result, we obtain the \textit{VI}-curves of the
sample with PAPS and study the effects of temperature variation,
thermal fluctuations, and positional disorder of the pinning sites
on these curves. In Section IV an analytical criterion for the
development of the filamentary instability in superconductors with
an NDR of S-type is derived. In Section V we analyze the effect of
the interplay between S-type and N-type NDR on the flux flow in
superconductors with a PAPS. We argue that the coexistence of the
NDR of {\it both} types can give rise to macroscopic non-uniform
self-organized dynamical structures in the flux flow regime.

\section{Model}

\begin{figure}[btp]
\begin{center}
\includegraphics*[width=8.0cm]{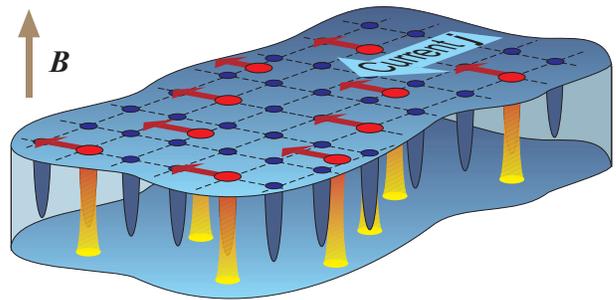}
\end{center}
\caption{ The model: vortices driven by a Lorentz
force produced by an applied current in a superconductor with
square array of pinning sites. The period of the pinning array is
$a = 2\lambda_{0}$. The pinning sites are shown by dark blue (dark
grey) parabolic bars and by dark blue (dark grey) dots on the
upper surface. The black dashed lines connecting the dots on the
top surface are a guide to the eye. The vortices are shown by
red-to-yellow (grey-to-light grey) tubes and by red (grey) larger
spots on the top surface. The direction of the applied current $j$
is indicated by a wide light blue (light grey) arrow, and the
Lorentz force acting on the vortices $F_{d}$ is shown by small red
(grey) arrows. The direction of the external applied magnetic
field $B$ is shown by the brown (grey) arrow. }
\end{figure}

We describe the flux motion in a three-dimensional (3D)
superconducting slab, infinite in the $xy$-plane, using a 2D model
(assuming no changes in the $z$-direction). This approach has also
been used in the past, e.g., in
Refs.~\onlinecite{fn1,weprl,md01,md03Z}. We consider a sample with
a square array of $N_p$ pinning sites interacting with $N_v$
vortices related to the magnetic field by $B=N_v\Phi_0$.  The
magnetic field is perpendicular to the slab (see Fig.~1). The
period of the regular array is $a = 2\lambda_{0}$ and we focus on
the case when the magnetic field $B$ is slightly higher than the
first matching field $B_\phi$, that is $N_v>N_p$. The vortices are
driven by the Lorentz force, $F_d=j\Phi_0/c$, produced by the
current flowing in the $x$ direction (see Fig.~1). Thus, the
horizontal axis of figures 2, 3, 4, and 5, refer to the driving
force $F_d$ or driving current $j$, since these are proportional
to each other. The overdamped motion of the $i$th vortex is
described by the equation
\begin{equation} \label{1}
\eta
\mathbf{v}_i=\mathbf{F}_i^{vv}+\mathbf{F}_i^{vp}+\mathbf{F}_{i}^{T}+\mathbf{F}_d,
\end{equation}
where $\mathbf{v}_i$ is the velocity of $i$th vortex,
$\eta=\sigma_nH_{c2}(T)\Phi_0/c^2$ is the flux flow viscosity,
$\sigma_n$ is the normal conductivity, and $H_{c2}$ is the upper
critical field. $\mathbf{F}_i^{vv}$ is the force per unit length
acting on the $i$th vortex due to the interaction with other
vortices. The force per unit vortex length $\mathbf{F}_i^{vp}$
describes the interaction of the $i$th vortex with the pinning
array. The term $\mathbf{F}_{i}^{T}$ arises due to the thermal
fluctuation contribution to the force. As in standard approaches
$\mathbf{F}_{i}^{T}(t)$ is a random function of time $t$, obeying
the correlation relations
\begin{equation} \label{2}
\langle F_{i}^{T}(t) \rangle_{t} = 0
\end{equation}
and
\begin{equation} \label{3}
\langle F_{i}^{T}(t)F_{j}^{T}(t^{\prime}) \rangle_{t} = 2 \eta
k_{B} T \delta_{ij} \delta(t-t^{\prime}),
\end{equation}
where $k_B$ is the Boltzmann constant, $\langle ... \rangle_{t}$
denotes a time average, $\delta_{ij}$ is the Kronecker symbol, and
$\delta(t)$ delta-function.

We describe the vortex-vortex interaction by the usual expression
for Abrikosov vortices
\begin{equation} \label{4}
\mathbf{F}_i^{vv}=\left( \frac{\Phi_0^2}{8\pi^2\lambda^3(T)}
\right) \sum_{j=1}^{N_v} \, K_1 \! \left(
\frac{|\mathbf{r}_i-\mathbf{r}_j|}{\lambda(T)} \right)
\mathbf{\widehat{r}}_{ij},
\end{equation}
where $\lambda$ is the magnetic field penetration depth, $K_1$ is
the first order modified Bessel function, the summation is
performed over the positions $\mathbf{r}_j$ of $N_v$ vortices in
the sample, and
$\mathbf{\widehat{r}}_{ij}=(\mathbf{r}_i-\mathbf{r}_j)/|\mathbf{r}_i-\mathbf{r}_j|$
is a unit vector in the direction of the force acting between the
$i$th and $j$th vortices.

The $N_p$ pinning sites (narrow indentations or ``blind holes''
which can accomodate a maximum of one vortex, for the vortex
densities used in our calculations) are located at positions
$\mathbf{r}_k^{(p)}$. The pinning potentials are approximated by
parabolic wells. Then, the pinning force per unit length acting on
the $i$th vortex can be written in the form
\begin{equation} \label{5}
\mathbf{F}_i^{vp}= \left( \frac{F_p(T)}{r_p} \right)
\sum_{k=1}^{N_p}|\mathbf{r}_i-\mathbf{r}_k^{(p)}| \, \Theta \!
\left( \frac{r_p-|\mathbf{r}_i-\mathbf{r}_k^{(p)}|}{\lambda_0}
\right) \mathbf{\widehat{r}}_{ik}^{(p)},
\end{equation}
where $r_p$ is the size of the elementary pinning potential well,
$F_p(T)$ is the maximum pinning force, $\Theta$ is the Heaviside
step function, $\lambda_0=\lambda(T=0)$, and
$\mathbf{\widehat{r}}_{ik}^{(p)}=(\mathbf{r}_i-\mathbf{r}_k^{(p)})/|\mathbf{r}_i-\mathbf{r}_k^{(p)}|$
is the unit vector in the direction of the elementary pinning
force. In what follows, we estimate the maximum pinning force as
\begin{equation}\label{FpT}
F_p(T)=\frac{H_c^2(T)\xi^2(T)}{r_p}\,,
\end{equation}
where $\xi(T)$ is the coherence length.

The temperature dependence of the values entering our model is
found using the Ginzburg-Landau approach. Therefore,
$H_{c2}(T)=\Phi_0/2\pi\xi^2(T)$. The temperature dependence of the
penetration depth $\lambda$ is approximated as
\begin{equation} \label{6}
\lambda(T)=\lambda_0 \left( 1-\frac{T^2}{T_c^2} \right) ^{-1/2}.
\end{equation}
We assume that the Ginzburg-Landau ratio $\kappa_{GL}=\lambda/\xi$
is independent of temperature. In this case, for the temperature
dependence of the maximum pinning force, we have
\begin{equation} \label{7}
F_p(T)=F_{p0} \left( 1- \frac{T^2}{T_c^2} \right),
\end{equation}
where $F_{p0}=F_{p0}(T=0)$. We also assume that the normal
conductivity $\sigma_n$ is temperature independent.

We simulate Eqs.~(\ref{1})--(\ref{5}) using the molecular dynamics
technique. Below we present the results for rectangular cells with
size $18\times12$~$\lambda_0^2$. Periodic boundary conditions are
imposed at the cell boundaries. First, we should prepare an
initial state of our system. For this purpose we assume that the
initial temperature of the system is high and that the vortex
structure is in a liquid unpinned state. Then, we slowly decrease
the temperature down to $T = 0$ and vortices are captured by the
pinning sites. When cooling down, vortices adjust themselves to
minimize their energy, simulating field-cooled experiments.
Starting from this initial state, we increase the driving current
and compute the average vortex velocity $\bar{v}(j)$, which is
determined as
\begin{equation}\label{8}
\bar{v}(j)=N_v^{-1}\sum_i \mathbf{v}_i\cdot \hat{\mathbf{x}},
\end{equation}
where $\hat{\mathbf{x}}$ is the unit vector in the $x$ direction.

The equilibrium temperature distribution in the sample can be
found by solving the heat equation with Joule heating. The average
power of this heating per unit volume due to vortex motion is
$jE=j\bar{v}B/c$. The heat flux, $q$, removed from the sample
boundaries by the external coolant is described by the usual
linear Kapitza law, $q=Sh_0(T-T_0)$, where $S$ is the sample
surface, $h_0$ is the heat transfer coefficient, and $T_0$ the
ambient temperature. To simplify the procedure, we assume that the
heat conductivity of the sample, $\kappa$, is large, $\kappa\gg
h_0w$, where $w$ is the sample thickness. Under such a condition,
the temperature in the sample is uniform and it can be found from
the heat balance equation~\cite{GMR}:
\begin{equation}
h_0S\left(T-T_0\right)=\frac{\bar{v}}{c}jBV, \label{hs}
\end{equation}
where $V$ is the sample volume. Further, we shall
assume that $T_0\ll T_c$ and neglect $T_{0}$.

We now introduce dimensionless variables. The dimensionless
current (which is equal to the dimensionless driving force $f_d$),
dimensionless average vortex velocity $V_x$, and dimensionless
magnetic field induction $b$, are given by
\begin{equation}\label{dim}
f_d=\frac{j}{j_0},\,\,\,V_x=\frac{\bar{v}}{v_0},\,\,\,b=\frac{B}{B_{\phi}}.
\end{equation}
The normalization values $j_0$ and $v_0$ are defined by
\begin{equation}
j_0 = \frac{ c\Phi_0 }{ 8\pi^2\lambda_0^3 }, \ \ \ v_0 = \frac{
c^2 }{ 4\pi\kappa^2_{GL}\sigma_n\lambda_0 }. \label{v0j0}
\end{equation}
In dimensionless units the heat balance equation (\ref{hs})
relates the temperature and the dimensionless driving force as
\[
\frac{T}{T_c}=K_{\textrm{th}} \, V_x \, f_d \, b,
\]
where
\begin{equation}
K_{\textrm{th}} = \frac{ j_0 \, v_0 \, B_{\phi} \, V }{ c \, h_0
\, T_c \, S } \label{v0j0}
\end{equation}
is the ratio of the characteristic heat release to heat removal.
In our calculations we used the values of the parameters
characteristic of high temperature superconductors:
$\lambda_0=2000$~\AA, $\kappa_{GL}=100$,
$\sigma_n=10^{16}$~s$^{-1}$, $V/S=1000$~\AA, $T_c=90$~K,
$B_{\phi}=500$~G, and $h_0=1$~W/cm$^2$K. In this case, we find
that $K_{\textrm{th}}=0.05-0.06$ and $F_{p0}$ is of the order of
$F_0=\Phi_0j_0/c$. In the simulations, we used $K_{th}=0.0525$ and
$F_{p0}=2F_0$ .

\section{Simulation results}

\subsection{Effect of heating: {\it VI}-curve of N- and S-type}

The calculated dependence of $\bar{v}$ versus the current density
$j$ in the absence of the Joule heating is shown in Fig.~2a, in
dimensionless units. The shape of this curve is similar to that as
found in Ref.~\onlinecite{fn1} and \onlinecite{weprl}: there exist
five dynamical vortex phases described in the introduction and a
pronounced hysteresis for increasing and decreasing current
regimes.

A significant effect of the heating is observed if the current
density exceeds some threshold value, $j \gtrsim 3.75$ for the
case shown in Fig.~2b. In particular, the jump from phase II to
phase III becomes larger, and an abrupt transition occurs between
regimes IV and V, compared to the non-heating case shown in
Fig.~2a. The most important feature related to Joule heating in
the high current range is the appearance of hysteresis in some
regions of phases IV and V. For decreasing current, the overheated
vortex lattice keeps moving as a whole at lower currents than the
``cold'' one (for increasing $j$). The transition part of the
$\bar{v}(j)$ curve from phase IV to phase V is shown by the dashed
line in Fig.~2b. This part of the \textit{VI}-curve is unstable
for a given drive force and could be only found for a fixed
voltage~\cite{GM}.

As a result, we obtain a new complex NDR of a hybrid nature with
\textit{both} $N$- and $S$-type instabilities, which is very
unusual for any media, especially for superconductors. Each type
of NDR is characterized by its specific
instabilities~\cite{sch,sha,MR,GM,GMR,cap,mikhailovskii}. Thus,
the obtained \textit{VI}-curve is characterized by {\it two kinds
of instabilities}. For example, if the current density exceeds the
value $j\approx 3.5$ (point A in Fig.~2b), the uniform current
flow becomes unstable and a filamentary instability~\cite{sch,sha}
occurs. Due to this instability, the current flow breaks into
filaments with different supercurrent density, some with lower
current $j_B$ (state B) and others with higher current $j_C$
(state C). However, the state C is, in its turn, unstable and
decays. The corresponding stable states are on the lower (E) and
on the upper (D) \textit{VI}-curve branches. That is, the filament
breaks into domains with higher and lower value of vortex flow
speed $\bar{v}$; in other words, with higher and lower electric
field. The stability and evolution of such a complicated structure
is an open question (see also Section V).

\begin{figure}[btp]
\begin{center}
\includegraphics*[width=7.5cm]{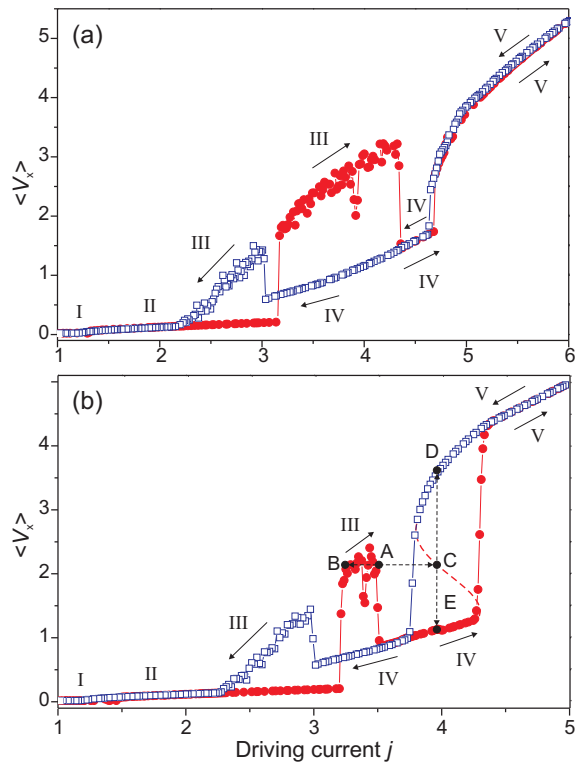}
\end{center}
\vspace{-0.5cm} \caption{ The average vortex
velocity $\bar{v} \propto E$ versus current $j$ for $B/B_{\phi} =
1.074$, $r_{p} = 0.21\lambda_{0}$ and $F_{p0}/F_{0} = 2$ for
increasing (shown by red (black) solid circles) and for decreasing
(blue open squares) $j$ (see, Ref.~\cite{weprl}).
(a) No heating effect is taken into account. The regions
corresponding to different phases are indicated by the roman
numerals from I to V (as in Ref.~\cite{fn1,weprl}).
(b) The sample heats up due to vortex motion. For small values of
the drive $j$ (up to $j \sim 3$) the effect of heating is
negligible. For $j \sim 3.2$, a jump from phase II to phase III
occurs. As a result of the heating, a transition occurs abruptly
at $j \sim 4.3$, from regime IV to regime V, where the vortex
lattice is entirely unpinned and moves as a whole. A hysteresis
now appears in region V: when decreasing the current $j$ down to
the value at which the jump from phase IV to phase V occurred,
when the driving increased, the overheated vortex lattice keeps
moving as a whole. As a result, we obtain a complicated $N$- and
$S$-type {\it VI}-curve. State A is unstable and the sample
divides into filaments in states B and C. State C is also
unstable. The corresponding stable states are on the lower (point
E) and on the upper (point D) {\it VI}-curve branches. Point $j
\sim 4.5$ corresponds to the normal transition and at $T>T_c$ we
have the usual Ohmic conductivity.}
\end{figure}

\begin{figure}[btp]
\begin{center}
\hspace{-0.5cm}
\includegraphics*[width=8.0cm]{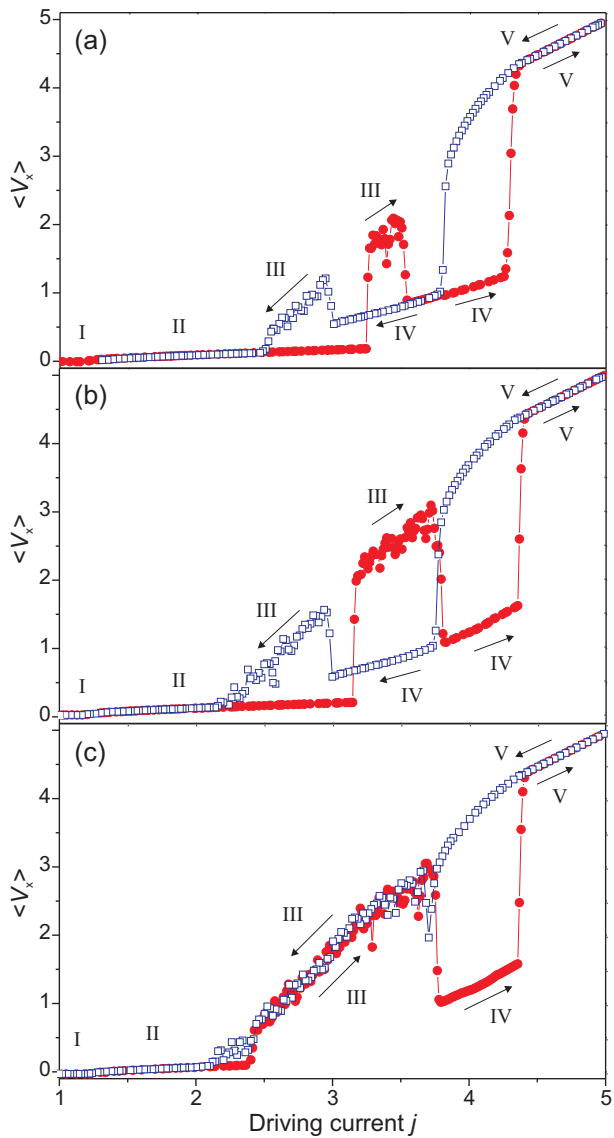}
\end{center}
\vspace{-0.5cm} \caption{ The average vortex
velocity $\bar{v}(j)$ for $B/B_{\phi} = 1.074$, $F_{p0}/F_{0} =
2.0$ for increasing (red/black solid circles) and for decreasing
(blue open squares) $j$ and different radii of the pinning sites:
(a) $r_{p} = 0.23\lambda_{0}$; (b) $r_{p} = 0.20\lambda_{0}$; (c)
$r_{p} = 0.19\lambda_{0}$. The function $\bar{v}(j)$ only slightly
changes for radii from $r_{p} = 0.20\lambda_{0}$ to
$0.25\lambda_{0}$. For smaller $r_{p}$, phase IV in the reverse
branch disappears (c) since the overheated vortex lattice cannot
adjust itself to the pinning array.}
\end{figure}

In Fig.~3, the average vortex velocity $\bar{v}(j)$ is shown for
different radii of the pinning sites. The shape of the function
$\bar{v}(j)$ only slightly changes for the radii in the range
$r_{p} = 0.20\lambda_{0}$ to $0.25\lambda_{0}$ (Figs.~3a, b).
However, for radii smaller than a certain value, phase IV in the
reverse branch (i.e., when decreasing the driving current $j$)
disappears (Fig.~3c) since the overheated vortex lattice cannot
adjust itself to the pinning array and turns to the disordered
motion in phase III.

\subsection{Effect of disorder}

Let us study the effects of disorder on the NDR. Small disorder
can be effectively introduced to the system by increasing the
radius, $r_{p}$, of the pinning sites. So vortices acquire an
additional degree of freedom and can move inside the pinning
sites. The function $\bar{v}(j)$ is shown in Fig.~4 for larger
pinning site radii, $r_{p} = 0.35\lambda_{0}$ (Fig.~4a), and
$r_{p} = 0.45\lambda_{0}$ (Fig.~4b). In case of larger radii,
phase III disappears, and the motion of interstitial vortices
(phase II) transforms directly to the 1D incommensurate vortex
motion (phase IV) (Fig.~4a). However, {\it the robust hysteresis
related to heating remains}. For larger radii of the pinning
sites, phase II, related to the motion of interstitial vortices,
disappears since all the vortices are pinned for weak enough
drives (Fig.~4b).

\begin{figure}[btp]
\begin{center}
\hspace{-0.5cm}
\includegraphics*[width=8.0cm]{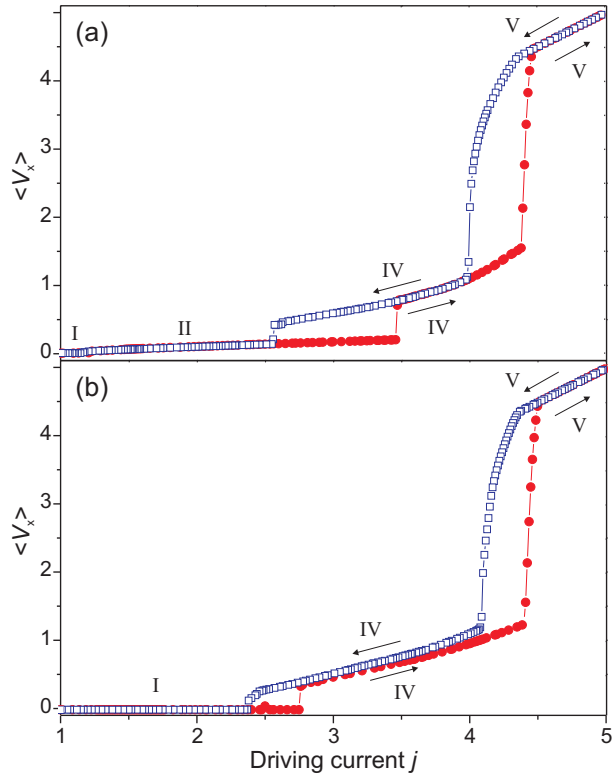}
\end{center}
\vspace{-0.5cm} \caption{ The average vortex
velocity $\bar{v}(j)$ for increasing (red (black) solid circles)
and for decreasing (blue open squares) $j$ for large pinning sites
radii: (a) $r_{p} = 0.35\lambda_{0}$; (b) $r_{p} =
0.45\lambda_{0}$. Other parameters are the same as in Fig.~2. An
increase of $r_{p}$ increases the disorder in the system because
vortices can then move inside the pinning sites. Phase III
disappears in (a). However, the hysteresis related to the heating
remains. For larger $r_{p}$ (b), phase II disappears since all the
vortices are pinned.}
\end{figure}

To model disorder related to a distortion of the regular PAPS, we
introduced small random displacements for each pinning site.
Specifically, for the displacement of each pinning site we
consider a random angle $\alpha_{\rm ran}$ ($0 < \alpha_{\rm ran}
< 2\pi$) and a random radius $r_{\rm ran}$ ($0 \leq r_{\rm ran}
\leq r_{\rm ran}^{\rm max}$, measured in units of $a/2$, where $a$
is a period of the (regular) pinning array) for each pinning
displacement.

\begin{figure}[btp]
\begin{center}
\hspace{-0.5cm}
\includegraphics*[width=8.0cm]{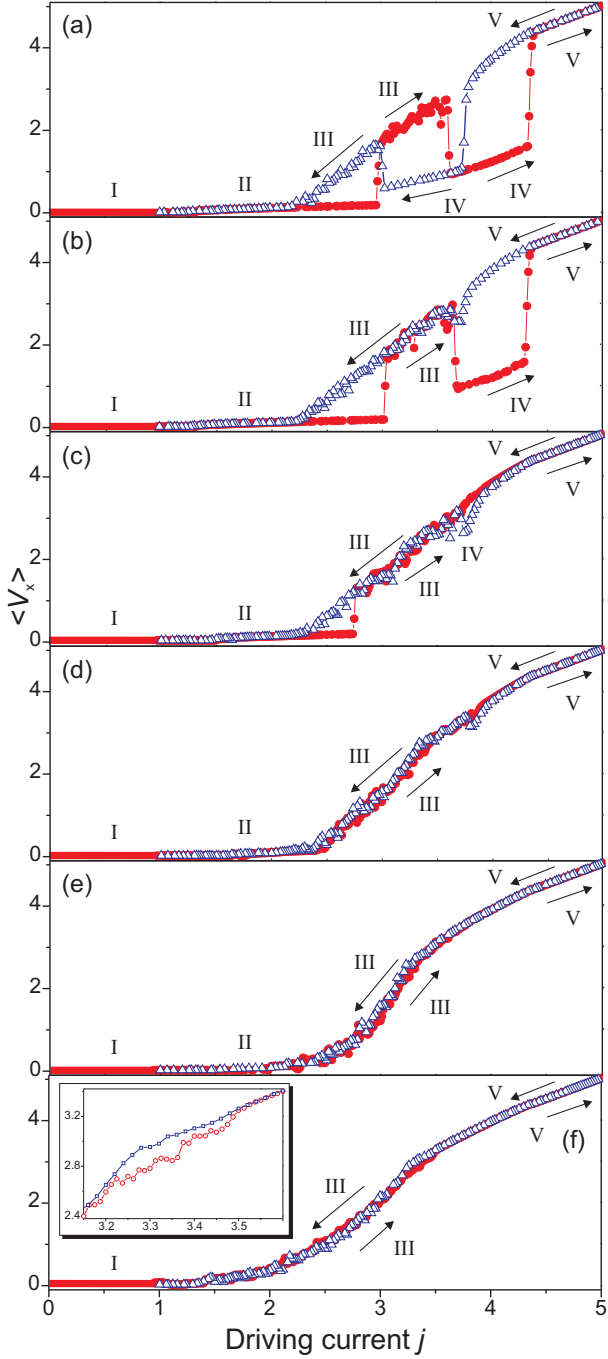}
\end{center}
\vspace{-0.5cm} \caption{ \setlength{\baselineskip}{12pt minus3pt}
The average vortex velocity $\bar{v}$ as a function
of the driving current $j$ for $B/B_{\phi} = 1.074$, $r_{p} =
0.2\lambda_{0}$ and $F_{p0}/F_{0} = 2.0$ for increasing (red
(black) solid circles) and for decreasing (blue open triangles)
$j$ for different amounts of disorder (displacement of the centers
of the pinning sites from their regular positions) in the system:
(a) $d_{\rm ran} = 0.01(a/2)$; (b) $d_{\rm ran} = 0.05(a/2)$; (c)
$d_{\rm ran} = 0.1(a/2)$; (d) $d_{\rm ran} = 0.2(a/2)$; (e)
$d_{\rm ran} = 0.5(a/2)$; (f) $d_{\rm ran} = a/2$. The function
$\bar{v}(j)$ does not appreciably change for small amount of
disorder (a). For $d_{\rm ran} > 0.05(a/2)$ (b), phase IV
disappears in the reverse branch. For $d_{\rm ran} > 0.1(a/2)$
(c), phase IV is lost in both branches; only some reminiscent
features remain, which disappear for larger $d_{\rm ran}$ (d, e).
At maximal disorder $d_{\rm ran} = a/2$ (f), only phases I, III
and V remain. However, there is a weak hysteresis related to
heating (inset to (f)).}
\end{figure}

The corresponding values of $\bar{v}(j)$ are shown in Fig.~5 for
different amounts of disorder. Note that small disorder does
\textit{not} appreciably influence $\bar{v}(j)$ (Fig.~5a).
However, for $r_{\rm ran}^{\rm max} > 0.05(a/2)$ (Fig.~5b), phase
IV disappears in the reverse branch of $\bar{v}(j)$. For $r_{\rm
ran}^{\rm max} > 0.1(a/2)$ (Fig.~5c), phase IV is lost in both
branches; only some reminiscent features remain, which disappear
for larger $d_{\rm ran}^{\rm max}$ (Figs.~5d, e). Finally, at full
disorder $r_{\rm ran}^{\rm max} = a/2$ (Fig.~5f), only phases I,
III and V remain. However, even in this case there is a weak
hysteresis related to heating (inset to Fig.~5f), observed in
experiments with random pinning \cite{GMR}.

\section{Filamentary instability}

It is well-known that the uniform current and electric field
distributions are \textit{unstable} under definite conditions if
the sample \textit{VI}-curve  has parts with
NDR~\cite{sch,sha,MR,GM,GMR,cap,mikhailovskii}. Let us assume that
the sample with the \textit{VI}-curve  shown in Fig.~2b is in a
current-biased regime. If the driving current exceeds the value $j
\approx 3.75$, the uniform current flow with the current density
$j$ becomes unstable with respect to the so-called filamentary
instability~\cite{sch,sha}. Thus, the current flow in the sample
breaks up into stripes or filaments with two different alternating
current densities. This process is illustrated in Fig.~1b: the
sample in state A with current density $j_A$ breaks into filaments
with lower $j_B$ (state B) and higher $j_C$ (state C) current
densities.

\begin{figure}[btp]
\begin{center}
\vspace*{-7.0cm}
\hspace{-1.0cm}
\includegraphics*[width=9.5cm]{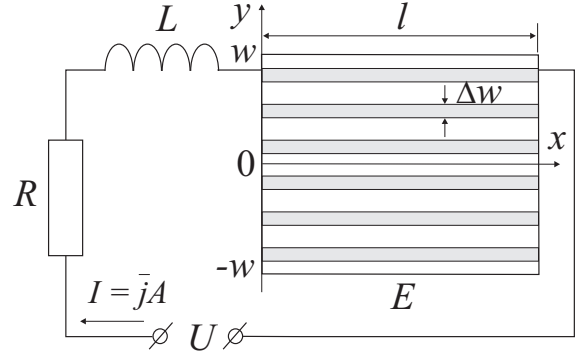}
\end{center}
\vspace{-1.5cm}
\caption{
The electrical circuit. The current
filaments are shown schematically by light grey lines \cite{weprl};
$l$ is the sample length, $w$ is the sample half-width, $\Delta w$ is the
filament width, the boundary conditions are stated at the sample
edges $y=\pm w$.
}
\end{figure}

To study the process in more detail, we consider a sample
connected to a standard electrical circuit, Fig.~6. The circuit
equation is
\begin{equation}\label{A1}
L \frac{\partial I}{\partial t} + RI + lE = U,
\end{equation}
where $I$ is the current in the circuit, $L$ and $R$ are the
circuit inductance and resistance, $l$ is the sample length, $E$
is the electric field in the sample, $U$ is the voltage at the
circuit terminals, which is assumed to be constant, and $j=I/A$,
where $A$ is the sample cross-section. Using Eq.~(\ref{A1}) and
Maxwell equations we can write the equations describing the
development of the small perturbations of electromagnetic field
$\delta \mathbf{E}$, $\delta \mathbf{B}$, and current density
$\delta \mathbf{j}$ in the form
\begin{eqnarray}\label{A2}
 \nonumber LA \; \frac{\partial}{\partial t} (\delta \bar{j}) + RA \; (\delta \bar{j}) + l \delta \! \bar{E}=0, \\
  \nabla \times \delta \mathbf{E} = -\frac{1}{c} \frac{\partial (\delta \mathbf{B})}{\partial t}, \\
 \nonumber \nabla \times \delta \mathbf{B} = \frac{4\pi}{c} (\delta \mathbf{j}).
\end{eqnarray}
Here
\[
\delta \mathbf{j} = \frac{\partial j}{\partial E} (\delta
\mathbf{E}) + \frac{\partial j}{\partial B} (\delta \mathbf{B})\,,
\]
all quantities are assumed averaged over a volume including a
large number of vortices, $\delta \bar{j}$ and $\delta \bar{E}$
denote the average values over the sample cross-section. Below we
assume for simplicity that the sample has zero demagnetization
factor.

We shall seek the solution to Eqs.~(\ref{A2}) in the standard
form: $\delta E,\,\delta B \propto \exp(\lambda t/t_0)$; where
$\lambda$ is the value to be found, and $t_0=L/R$ is the circuit
relaxation time.

An instability develops if $Re\ (\lambda) > 0$. In general we
should add to Eqs.~(\ref{A2}) the equation for the small
temperature perturbations but here we study the filamentary
instability for which the temperature rise is not of crucial
importance. We also neglect the self-field effect and assume that
the background magnetic field in the sample is uniform. To find
the filamentary instability criterion we can consider the
perturbations depending on the $y$ coordinate only~\cite{sch,sha}.
In such a geometry, the perturbation of the electric field has
only the $x$ component, while $\delta \mathbf{B}$ has only the $z$
component. Using $\delta B=ct_0\delta E'/\lambda w$, we find from
Eqs.~(\ref{A2})
\begin{eqnarray}
\label{A3} (\lambda+1)\delta \bar{j}+\rho_c^{-1}\delta \bar{E} &=& 0, \\
\label{A4}  \delta E''-\beta \delta E'-\frac{\lambda t_s}{t_0}
\delta E &=& 0,
\end{eqnarray}
where prime means differentiation over the dimensionless
coordinate $y/w$, $w$ is the sample half-width, $ \rho_c=RA/l$,
and
\[
t_s=\frac{4\pi w^2}{c^2}\frac{\partial j}{\partial
E}\,,\quad\,\beta=\frac{4\pi w}{c}\frac{\partial j}{\partial B}
\]
are the sample \textit{VI}-curve  parameters, which are positive
or negative depending on the relevant part of the
\textit{VI}-curve at given background fields $E$ and $B$. Note
that the value of $|t_s|$ is the characteristic time of the
magnetic field relaxation in the sample.

The differential equation (\ref{A4}) requires two boundary
conditions. We obtain the first one assuming that the applied
magnetic field $B$ is constant. This means that $\delta
B(1)=-\delta B(-1)$ or $\delta E'(1)=-\delta E'(-1)$. From the
Maxwell equations (\ref{A2}) we get $\delta B(1)-\delta B(-1)=8\pi
w \delta\bar{j}/c$. Using these $\delta B$'s, and substituting
$\delta \bar{j}$ from Eq.~(\ref{A3}), we obtain the second
boundary condition in the form
$$
\delta E'(1) = - \frac {\gamma \lambda \delta
\bar{E}}{(\lambda+1)},
$$
where
$$
\gamma = \frac{4 \pi w^2 l}{c^2 A L}.
$$

The solution of Eq.~(\ref{A4}) reads
\begin{equation}\label{A8}
\delta E=C_1\exp(p_1y/w)+C_2\exp(p_2y/w),
\end{equation}
where $C_i$ are constants and
\[
p_{1,2}=\frac{\beta}{2}\pm \sqrt{\frac{\beta^2}{4}+\frac{\lambda
t_s}{t_0}}.
\]
Substituting Eq.~(\ref{A8}) to the boundary conditions we obtain a
set of uniform linear equations for the constants $C_{1,2}$. The
non-trivial solution of this equation set exists if its
determinant is zero. Thus we find the equation for the eigenvalue
spectrum $\lambda$ in the form
\[
p_1\left[p_2+\frac{\gamma\lambda}{(\lambda+1)p_2}\right]\cosh
p_1\sinh p_2=
\]
\begin{equation}\label{A9}
p_2\left[p_1+\frac{\gamma\lambda}{(\lambda+1)p_1}\right]\sinh
p_1\cosh p_2.
\end{equation}

In the simplest case of small $|\partial j/\partial B|$, when
\[
p_1=-p_2=\sqrt{\lambda t_s/t_0},
\]
the solution of Eq.~(\ref{A9}) can be readily found explicitly
with an accuracy up to $|\beta^2|$
\begin{equation}
\label{A10} \lambda \, = \, -1-\frac{\gamma t_0}{t_s} \, = \,
-1-\rho_c^{-1}\frac{\partial E}{\partial j}.
\end{equation}
It follows from the last relation that the instability occurs only
at the \textit{VI}-curve branch with NDR when
\[
t_s\propto \frac{\partial j}{\partial E}<0
\]
and, moreover, the drop of the voltage should be large enough,
\begin{equation}\label{crit}
\left|\frac{\partial E}{\partial j}\right|>\rho_c\,.
\end{equation}
In this case $p_{1,2}$ are purely imaginary numbers and the
solution of Eqs.~(14) is periodic in the $y$-direction. The
characteristic spatial period of the arising current filament
structure is of the order of
\begin{equation}
\label{A12} \Delta w \, \propto \, \frac{w}{|p_1(\lambda)|} \, =
\, \frac{w}{\sqrt{|\gamma+t_s/t_0|}}.
\end{equation}
In a very unstable state, $| \gamma t_s/t_0 | \gg 1$, the filament
width is small, $\Delta w \ll w$. The characteristic instability
build-up time becomes $t_0/\lambda$. Thus, a sample with an S-type
NDR in \textit{VI}-curve divides itself into small filaments with
different current densities (in different dynamic flux flow phases
III and IV) if the resistance and inductance of the external
circuit are restricted by inequalities
\begin{equation}\label{A13}
R\ll \frac{A}{l}\left|\frac{\partial E}{\partial j}\right|\,,\quad
L\ll 4\pi l/c^2A\,.
\end{equation}

The obtained results are valid if
\[
\left|\rho_c^{-1}\frac{\partial E}{\partial J}\right|\gg
\left(\frac{4\pi w}{c}\frac{\partial j}{\partial B}\right)^2.
\]
According to Eq.~(\ref{crit}), the left hand side of the last
inequality should be higher than unity, while the right hand side
is much smaller than 1 for the parameter range studied in the
previous sections if the sample half-width $w<1$~mm. Note,
however, that taking into account the magnetic field dependence of
the \textit{VI}-curve  gives rise to some increase in its
stability.

\section{Interplay between N-type and S-type instabilities}

In the stationary inhomogeneous state that arises after the
development of the filamentary instability, the electric field
should be uniform over the sample. The part, $p$, of the filament
with the higher current $j_C$ and the part, $1-p$, with the lower
current $j_B$ (see Fig.~2b) are defined by an evident condition
$j_A=pj_C+(1-p)j_B$, that is,
$$
p = \frac{(j_C-j_A)}{(j_C-j_B)}.
$$
For the case shown in Fig.~1b we find an estimate $p\approx 0.6$.

As shown in Fig.~4, sufficiently high disorder destroys the NDR of
S-type in the \textit{VI}-curve, but the NDR of N-type can still
exist in this case due to sample heating. Such an unstable regime
has been thoroughly studied for superconductors~\cite{MR,GM,GMR},
and we do not discuss it here.

A richer dynamics can be observed if the \textit{VI}-curve has NDR
parts of \textit{both} N- and S-types, Figs.~2b and 3. In this
case the filaments with higher current density $j_C$ are unstable
if the system is far from the voltage-biased
regime~\cite{sch,sha}. The instability of the filament with N-type
\textit{VI}-curve should switch the filaments into state D
(Fig.~2b) with high resistivity or to the formation of the domain
structure with higher, D, and lower, E, resistivity~\cite{GM}.
However, any possible decay of the unstable state C gives rise to
a non-uniform electric field distribution in the sample and, as a
result, to non-zero $\partial B/\partial t$. Thus, the state that
appears after the instability develops is not stationary but
rather a dynamic one.

\begin{figure}[btp]
\begin{center}
\hspace{-0.5cm}
\includegraphics*[width=8.5cm]{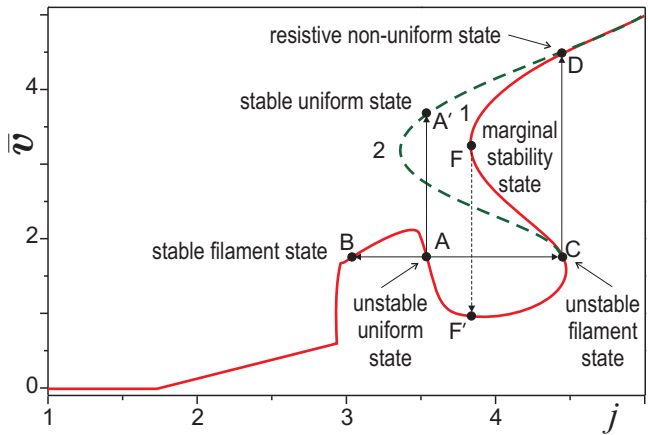}
\end{center}
\vspace{-0.5cm}
\caption{ Schematic \textit{VI}-curves (red (black)
solid line and green (dark grey) dashed line) of the
superconductor for two different values of the hysteresis due to
overheating \cite{weprl}. The more pronounced hysteresis loop
(shown by the green (dark grey) dashed line) corresponds to a
larger value of the characteristic heat release.}
\end{figure}

To clarify the situation, we consider two possible
\textit{VI}-curves shown in Fig.~7 (one red and another, labelled
by ''2'', with a green branch) and assume that the total current
value in the circuit is fixed, $I=Aj_A$. In a more general case,
after the decay of the unstable state C, the filaments with higher
current density will be overheated and transit to the higher
resistive state D. In this state D the flux lines move fast,
which, along with the temperature increase due to the thermal
conductivity, gives rise to the acceleration of the flux flow in
the lower-current filaments. As a result, the high-resistivity
overheated state moves from point D to a lower electric field
range, while the low-resistivity state (point B) moves to a higher
electric field range (point A). If we assume that the system has a
\textit{VI}-curve  of the type 2 (green (dark grey) dashed curve
in Fig.~7), then the high-current filaments in D move to the state
A$'$ with current density $j_A$ and a higher electric field than
in the initial state A. The filaments in point B move to point A
and jump to point A$'$. As a result, a new stable uniform state
with $j = j_{A}$ appears. However, if the \textit{VI}-curve has
the form shown by the red curve 1, the stable stationary point
with the current density $j_A$ does not exist. In this case, the
high electric field state moves from point D to the marginal
stability point F and then falls down to a lower branch of the
\textit{VI}-curve curve (point F$'$). In this state F$'$ the
electric field is lower than in the state B with lower current. As
result, the state moves from F to F$'$ and then to point A, which
is the only uniform state corresponding to the fixed current value
$j_A$. However, this state is unstable, and the cycle
$$
{\rm A} \rightarrow {\rm C} \rightarrow {\rm D} \rightarrow {\rm
F} \rightarrow {\rm F^{\prime}} \rightarrow {\rm A}
$$
is repeated (branch 1 in red in Fig.~7). Another branch (branch 2
with the green segment) of this cycle involves the stable filament
state B and is as follows:
$$
{\rm A} \rightarrow {\rm B} \rightarrow {\rm A}.
$$
Such a dynamic state has a non-stationary pattern of resistive
domains coexisting and intertwinned with current filaments.
The specific form of these patterns, and their dynamics can be
very complex, and requires investigations beyond the scope of
this study.

\vspace{1.0cm}

\section{Conclusions}

The influence of temperature on the dynamic phases and
current-voltage characteristic of superconductors with periodic
pinning array was investigated here. It is demonstrated that this
effect can change the \textit{VI}-curve drastically. For a range
of values of the pinning array parameters and heat transfer
characteristics it is possible to obtain the \textit{VI}-curves
with a negative differential resistivity of (i) either N- or
S-type, (ii) \textit{VI}-curves with \textit{both} types of NDR,
or \textit{VI}-curves \textit{without any} NDR parts. The uniform
flux flow is unstable if the \textit{VI}-curve  has a part with
NDR. The formation of resistive domain structures and/or
propagating the resistive state through the sample is a
characteristic of \textit{VI}-curves of N-type, while a
filamentary instability with sample regions (filaments) having
different current densities is a characteristic of
\textit{VI}-curves with NDR of S-type. Much more complex regimes
can be expected in the case of \textit{VI}-curves with NDR parts
of both types. In this case the possibility of arising dynamical
non-uniform regimes is argued.

\section*{ACKNOWLEDGMENT}

This work was supported in part by the National Security Agency
(NSA), Laboratory of Physical Sciences (LPS), Army Research
Office; and also supported by the US National Science Foundation
grant No.~EIA-0130383, JSPS-RFBR Grant No.~06-02-91200, RFBR Grant
No 06-02-16691, and RIKEN's President's funds. V.R.M. acknowledges
support through POD. S.S. acknowledges support from the Ministry
of Science, Culture and Sport of Japan via the Grant-in Aid for
Young Scientists No. 18740224 and EPSRC via No. EP/D072581/1.

\end{document}